\documentstyle[12pt]{article}
\begin{document}
\begin{titlepage}
\begin{flushright}
NTUA--96/00 \\
hep-ph/0010141
\end{flushright}
\vspace{1cm}

\begin{centering}
\vspace{.4in} {\Large {\bf Supersymmetry Breaking by Dimensional
Reduction \\
\vspace{.3cm}

over Coset Spaces}}\\
\vspace{1.5cm}

{\bf
P.~Manousselis}$^{a}$ and {\bf G.~Zoupanos}$^{b}$\\ \vspace{.2in}
Physics Department, National Technical University, \\ Zografou
Campus, 157 80, Athens, Greece.\\

\vspace{1.0in}

{\bf Abstract}\\

\vspace{.1in} We study the dimensional reduction of a
ten-dimensional supersymmetric $E_{8}$ gauge theory over
six-dimensional coset spaces. We find that the coset space
dimensional reduction over a symmetric coset space leaves the four
dimensional gauge theory without any track of the original
supersymmetry. On the contrary the dimensional reduction over a
non symmetric coset space leads to a softly broken supersymmetric
gauge theory in four dimensions. The $SO_{7}/SO_{6}$ and
$G_{2}/SU(3)$ are used as representative prototypes of symmetric
and non symmetric coset spaces respectively.
\end{centering}
\vspace{4.7cm}

\begin{flushleft}
$^{a}$e-mail address: pman@central.ntua.gr. Supported by
$\Gamma\Gamma$ET  grand 97E$\Lambda$/71.
\\ $^{b}$e-mail address:
George.Zoupanos@cern.ch. Partially supported by  the EU project
ERBFM-RXCT960090.
\end{flushleft}
\end{titlepage}
\section{Introduction}
Supersymmetry is an essential ingredient of most unification
attempts in the recent years. A celebrated aspect of supersymmetry
has been its role in providing a partial solution to the hierarchy
problem of GUTs \cite{Dim}, resulting from the non renormalization
theorem \cite{Wess}. Since the observed low energy world is
lacking of possessing supersymmetry in its known particle
spectrum, supersymmetry has to be broken. In fact, the
understanding of supersymmetry breaking is a fundamental open
question in all supersymmetric unification schemes and in
particular in superstring theories despite some attractive
proposals \cite{Sagnotti}. For phenomenological purposes the
supersymmetry breaking is parametrized by the corresponding soft
terms which are defined as those that do not spoil the ultraviolet
behaviour of supersymmetric theories \cite{Nilles}. In this spirit
the Minimal Supersymmetric Standard Model (MSSM) has been provided
with a soft supersymmetry breaking (SSB) sector parametrizing our
ignorance of a convincing mechanism.

Recently a lot of interest has been triggered by the possibility
that superstrings can be defined at a TeV scale  \cite{Anton}. The
string tension became an arbitrary parameter and can be anywhere
below the Planck scale and as low as TeV. The main advantage of
having the string tension at the TeV scale, besides the obvious
experimental interest, is that it offers an automatic protection
to the gauge hierarchy \cite{Anton}, alternative to low energy
supersymmetry \cite{Dim}, or dynamical electroweak symmetry
breaking \cite{Fahri},\cite{Marciano},\cite{Trianta}. However the
only vacua of string theory free of all pathologies are
supersymmetric. Then the original supersymmetry of the theory, not
being necessary in four dimensions, could be broken by the
dimensional reduction procedure.

The weakly coupled ten-dimensional $E_{8} \times E_{8}$
supersymmetric gauge theory is one of the few to posses the
advantage of anomaly freedom \cite{Green} and has been extensively
used in efforts to describe quantum gravity along with the
observed low energy interactions in the heterotic string framework
\cite{Theisen}. In addition its strong coupling limit  provides an
interesting example of the realization of the brane picture, i.e.
$E_{8}$ gauge fields and matter live on the two 10-dimensional
boundaries, while gravitons propagate in the eleven-dimensional
bulk \cite{Horava}.

Having a gauge theory defined in higher dimensions the obvious way
to dimensionally reduce it is to demand that the field dependence
on the extra coordinates is such that the Lagrangian is
independent of them. A crude way to achieve that is to disccard
the field dependence on the extra coordinates, while an elegant
one is to allow for a non trivial dependence on them, but impose
the condition that a symmetry transformation by an element of the
isometry group $S$ of $B$ corresponds to a gauge transformation.
Then the Lagrangian will be independent of the extra coordinates
just because it is gauge invariant. This is the basis of the Coset
Space Dimensional Reduction (CSDR) scheme \cite{Manton},
\cite{Review}, \cite{Kuby}, which assumes that $B$ is a compact
coset space, $S/R$. The requirement that transformations of the
fields under the action of the symmetry group of $S/R$ are
compensated by gauge transformations lead to certain constraints
on the fields. The solution of these constraints provides us with
the four-dimensional unconstrained fields as well as with the
gauge  invariance that remains in the theory after dimensional
reduction. It is interesting to note that the fields obtained
using the CSDR approach are the first terms in the expansion of
the D-dimensional fields in harmonics of the internal space $B$
and are massless after the first stage of the symmetry breaking.

The effective field theories resulting from compactification of higher
dimensional theories contain also towers of massive higher harmonics
(Kaluza-Klein) excitations, whose contributions at the quantum level
alter the behaviour of the running couplings from logarithmic to power
\cite{Taylor}. As a result the traditional picture of unification
of couplings may change drastically \cite{Dienes}.

Higher dimensional theories have also been studied at the quantum
level using the continuous Wilson renormalization group
\cite{Kubo} which can be formulated in any number of space-time
dimensions with results in agreement with the treatment involving
massive Kaluza-Klein excitations. In turn the CSDR approach can in
principle be exploited in the study of higher dimensional unified
field theories independently of reference to more general
frameworks like string theories.

In the present work we study the dimensional reduction of a
ten-dimensional supersymmetric $E_{8}$ gauge theory over a
symmetric and a non symmetric six-dimensional coset space. We show
that in the first case supersymmetry is broken completely by the
reduction procedure, while in the second a supesymmetric gauge
theory is obtained in four dimensions with a complete soft
supersymmetry breaking sector.
\section{The Coset Space Dimensional Reduction}
In the Coset Space Dimensional Reduction  (CSDR) scheme
\cite{Manton},\cite{Review},\cite{Kuby} one starts with a
Yang-Mills-Dirac Lagrangian with gauge group $G$ defined on a
 $D$-dimensional spacetime $M^{D}$, with metric $g^{MN}$, which is compactified to $ M^{4}
\times S/R$ with $S/R$ a reductive but in general non symmetric
coset space. The metric is assumed to have the form $$g^{MN}=
\left[\begin{array}{cc}\eta^{\mu\nu}&0\\0&-g^{ab}\end{array}
\right],$$ where $\eta^{\mu\nu}= diag(1,-1,-1,-1)$ and $g^{ab}$ is
the coset space metric. We can divide the generators of $S$, $
Q_{A}$ in two sets : the generators of $R$, $Q_{i}$ $(i=1,
\ldots,dimR)$, and the generators of $S/R$, $ Q_{a}$( $a=dimR+1
\ldots,dimS)$, and $dimS/R=dimS-dimR =d$. Then the commutation
relations for the generators of $S$ are the following :
\begin{eqnarray}
\left[ Q_{i},Q_{j} \right] &=& f_{ij}^k Q_{k},\nonumber \\
\left[ Q_{i},Q_{a} \right]&=& f_{ia}^{b}Q_{b},\nonumber\\
\left[ Q_{a},Q_{b} \right]&=& f_{ab}^{i}Q_{i}+f_{ab}^{c}Q_{c} .
\end{eqnarray}
When $S/R$ is symmetric, $f_{ab}^{c}=0$. Let us call the
coordinates of $M^{4} \times S/R$ space  $z^{M}=
(x^{m},y^{\alpha})$, where $\alpha$ is a curved index of the
coset,  $a$ is a tangent space index and $y$ defines an element of
$S$ which is a coset representative, $L(y)$. The vielbeins and
connection forms are defined through the Maurer-Cartan form which
takes values in the Lie algebra of $S$ :
\begin{equation}
L^{-1}(y)dL(y) = e^{A}_{\alpha}Q_{A}dy^{\alpha} .
\end{equation}
Doing a computation at the origin y=0 and using the fact that near
the origin $L(y) = exp(y^{a}Q_{a})$, we find that $ e^{a}_{\alpha}
= \delta^{a}_{\alpha}$ and $e^{i}_{\alpha} = 0$ . Now the group
$S$ acts as a symmetry group on the the extra coordinates (there
are $dimS$ Killing vectors which generate the isometries of $S/R$)
. The CSDR scheme demands that an $S$-transformation of the extra
$d$ coordinates is a gauge transformation of the fields that are
defined on $M^{4}\times S/R$,  thus a gauge invariant Lagrangian
written on this space is independent of the extra coordinates. Let
us see this in more detail.

Consider a $D$-dimensional Yang-Mills-Dirac theory with gauge group $G$
defined on a manifold $M^{D}$ which as stated will be compactified to
$M^{4}\times S/R$, $D=4+d$,  $d=dimS-dimR$:
\begin{equation}
A=\int d^{4}xd^{d}y\sqrt{-g}\Bigl[-\frac{1}{4}
Tr\left(F_{MN}F_{K\Lambda}\right)g^{MK}g^{N\Lambda}
+\frac{i}{2}\overline{\psi}\Gamma^{M}D_{M}\psi\Bigr] ,
\end{equation}
where
\begin{equation}
D_{M}= \partial_{M}-\theta_{M}-A_{M}
\end{equation}
with
\begin{equation}
\theta_{M}=\frac{1}{2}\theta_{MN\Lambda}\Sigma^{N\Lambda}
\end{equation}
the spin
connection of $M^{D}$, and
\begin{equation}
F_{MN} =\partial_{M}A_{N}-\partial_{N}A_{M}-\left[A_{M},A_{N}\right] ,
\end{equation}
where $M$,$N$ run over the $D$-dimensional space. The fields
$A_{M}$ and $\psi$ are, as explained, symmetric in the sense that
any transformation under symmetries of $S/R$  is compensated by
gauge transformations. The fermion fields can be in any
representation $F$ of $G$ unless a further symmetry such as
supersymmetry is required. So if $g(s)$, $f(s)$ are gauge
transformations in the adjoint and $F$ representations of $G$
corresponding to the transformation $s\in S$ acting on $S/R$, we
require
\begin{eqnarray}
A_{\mu}(x,y)&=&g(s)A_{\mu}(x,s^{-1}y)g^{-1}(s),\\
A_{\alpha}(x,y)&=&g(s)J_{\alpha}^{\beta}A_{\beta}(x,s^{-1}y)
g^{-1}(s)+g(s)\partial_{\alpha}g^{-1}(s),\\ \psi(x,y) &=&
f(s)\Omega\psi(x,s^{-1}y)f^{-1}(s),
\end{eqnarray}
where $J_{\beta}^{\alpha}$ is the Jacobian matrix for the
transformation s and $\Omega$ is the Jacobian matrix plus a local
Lorentz rotation in the tangent space which is needed for the
fermions when they transform in a curved space. These conditions
imply certain constraints that the $D$-dimensional fields have to
obey. The solution of these constraints will provide us with the
four-dimensional unconstrained fields as well as with the gauge
invariance that remains in the theory after dimensional reduction.
 From eq.(7) it follows that the components $A_{\mu}(x,y)$ of the initial
gauge field $A_{M}(x,y)$ become, after dimensional reduction, the
four-dimensional gauge fields and furthermore they are independent of
$y$. In addition one can find that they have to commute with the
elements
of the $R_{G}$ subgroup of $G$. Thus the four-dimensional gauge group
$H$ is
the centralizer of $R$ in $G$, $H=C_{G}(R_{G})$.

Similarly, from eq.(8) the $A_{\alpha}(x,y)$ components of
$A_{M}(x,y)$ denoted by $\phi_{\alpha}(x,y)$ from now on, become
scalars at four dimensions. These fields transform under $R$ as a
vector $v$, i.e.
\begin{eqnarray}
S &\supset& R \nonumber \\
adjS &=& adjR+v.
\end{eqnarray}
Moreover $\phi_{\alpha}(x,y)$ act as an intertwining operator
connecting induced representations of $R$ acting on $G$ and $S/R$.
This implies, exploiting Schur's lemma, that the transformation
properties of the fields $\phi_{\alpha}(x,y)$ under $H$ can be
found if we express the adjoint representation of $G$ in terms of
$R_{G} \times H$ :
\begin{eqnarray}
G &\supset& R_{G} \times H \nonumber \\
 adjG &=&(adjR,1)+(1,adjH)+\sum(r_{i},h_{i}).
\end{eqnarray}
Then if $v=\sum s_{i}$, where each $s_{i}$ is an irreducible
representation of $R$, there survives an $h_{i}$ multiplet for
every pair $(r_{i},s_{i})$, where $r_{i}$ and $s_{i}$ are
identical irreducible representations of $R$.

Turning next to the fermion fields \cite{Chapline},
\cite{Slansky}, \cite{Review} we see from eq.(9) that they act as
intertwining operators between induced representations acting on
$G$ and the tangent space of $S/R$, $SO(d)$. Proceeding along
similar lines as in the case of scalars to obtain the
representation of $H$ under which the four dimensional fermions
transform, we have to decompose the representation $F$ of the
initial gauge group in which the fermions are assigned under
$R_{G} \times H$, i.e.
\begin{equation}
F= \sum (t_{i},h_{i}),
\end{equation}
and the spinor of $SO(d)$ under $R$
\begin{equation}
\sigma_{d} = \sum \sigma_{j}.
\end{equation}
Then for each pair $t_{i}$ and $\sigma_{i}$, where $t_{i}$ and
$\sigma_{i}$ are identical irreducible representations there is an
$h_{i}$ multiplet of spinor fields in the four dimensional theory.
In order however  to obtain chiral fermions in the effective
theory we have to impose further requirements. We first impose the
Weyl condition in $D$ dimensions. In $D = 4n+2$ dimensions which
is the case at hand, the decomposition of the left handed, say
spinor under $SU(2) \times SU(2) \times SO(d)$ is
\begin{equation}
\sigma _{D} = (2,1,\sigma_{d}) + (1,2,\overline{\sigma}_{d}).
\end{equation}
So we have in this case the decompositions
\begin{equation}
\sigma_{d} = \sum \sigma_{k},~\overline{\sigma}_{d}= \sum
\overline{\sigma}_{k}.
\end{equation}

Let us start from a vector like representation $F$ for the
fermions. In this case each term $(t_{i},h_{i})$ in (12) will be
either self-conjugate or it will have a partner $(
\overline{t}_{i},\overline{h}_{i} )$. According to the rule
described in eqs.(12), (13) and considering $\sigma_{d}$ we will
have in four dimensions left-handed fermions transforming as $
f_{L} = \sum h^{L}_{k}$. It is important to notice that since
$\sigma_{d}$ is non self-conjugate $f_{L}$ is non self-conjugate
too. Similarly from $\overline{\sigma}_{d}$ we will obtain the
right handed representation $ f_{R}= \sum \overline{h}^{R}_{k}$
but as we have assumed that $F$ is vectorlike,
$\overline{h}^{R}_{k}\sim h^{L}_{k}$. Therefore there will appear
two sets of Weyl fermions with the same quantum numbers under $H$.
This is already a chiral theory, but still one can go further and
try to impose the Majorana condition in order to eliminate the
doubling of the fermionic spectrum. We should remark now that if
we had started with $F$ complex, we should have again a chiral
theory since in this case $\overline{h}^{R}_{k}$ is different from
$h^{L}_{k}$  $(\sigma_{d}$ non self-conjugate). Nevertheless
starting with $F$ vectorlike is much more appealing and will be
used in the following along with the Majorana condition. The
Majorana condition can be imposed in $D = 2,3,4+8n$ dimensions and
is given by $\psi = C(\overline\psi)^{T}$, where $C$ is the $D$
dimensional charge conjugation matrix. Majorana and Weyl
conditions are compatible in $D=4n+2$ dimensions. Then in our case
if we start with Weyl-Majorana spinors in $D=4n+2$ dimensions we
force $f_{R}$ to be the charge conjugate to $f_{L}$, thus arriving
in a theory with fermions only in $f_{L}$. Furthermore if $F$ is
to be real, then we have to have $D=2+8n$, while for $F$
pseudoreal $D=6+8n$.

Starting with an anomaly free theory in higher dimensions, in ref.
\cite{Witten} was given the condition that has to be fulfilled in
order to obtain anomaly free theories in four dimensions after
dimensional reduction. The condition restricts the allowed
embeddings of $R$ into $G$ \cite{Pilch}. For $G=E_{8}$ in ten
dimensions it has the form
\begin{equation}
l(G) = 60,
\end{equation}
where $l(G)$ is the sum over all indices of the $R$
representations appearing in the decomposition of the $248$
representation of $E_{8}$ under $ E_{8} \supset R \times H$. The
normalization is such that the vector representation in eq.(10)
which defines the embedding of $R$ into $SO(6)$, has index two.

 Next let us obtain the four dimensional effective action. Assuming that
the metric is block diagonal, taking into account all the
constraints and integrating out the extra coordinates we obtain in
four dimensions the following Lagrangian :
\begin{equation}
A=C \int d^{4}x \biggl( -\frac{1}{4} F^{t}_{\mu
\nu}{F^{t}}^{\mu\nu}+\frac{1}{2}(D_{\mu}\phi_{\alpha})^{t}
(D^{\mu}\phi^{\alpha})^{t}
+V(\phi)+\frac{i}{2}\overline{\psi}\Gamma^{\mu}D_{\mu}\psi-\frac{i}{2}
\overline{\psi}\Gamma^{a}D_{a}\psi\biggr),
\end{equation}
where $D_{\mu} = \partial_{\mu} - A_{\mu}$ and $D_{a}= \partial_{a}-
\theta_{a}-\phi_{a}$ with  $\theta_{a}=
\frac{1}{2}\theta_{abc}\Sigma^{bc}$ the connection of the coset space,
while $C$ is the volume of the coset space. The potential $V(\phi)$ is given
by:
\begin{equation}
V(\phi) = - \frac{1}{4} g^{ac}g^{bd}Tr( f _{ab}^{C}\phi_{C} -
[\phi_{a},\phi_{b}] ) (f_{cd}^{D}\phi_{D} - [\phi_{c},\phi_{d}] ) ,
\end{equation}
where, $A=1,\ldots,dimS$ and $f$ ' s are the structure constants appearing
in the commutators of the generators of the Lie algebra of S.

The expression (18) for $V(\phi)$ is only formal because
$\phi_{a}$ must satisfy the constraints coming from eq.(8),
\begin{equation}
f_{ai}^{D}\phi_{D} - [\phi_{a},\phi_{i}] = 0,
\end{equation}
where the $\phi_{i}$ generate $R_{G}$. These constraints imply
that some components $\phi_{a}$'s are zero, some are constants and
the rest can be identified with the genuine Higgs fields. When
$V(\phi)$ is expressed in terms of the unconstrained independent
Higgs fields, it remains a quartic polynomial which is invariant
under gauge transformations of the final gauge group $H$, and its
minimum determines the vacuum expectation values of the Higgs
fields \cite{Harnad}.

In the fermionic Lagrangian the first term is just the kinetic
term of the fermions, while the second is the Yukawa term
\cite{Kapetanakis}. We note that since $\psi$ is a Majorana-Weyl
spinor in ten dimensions the representation in which the fermions
are assigned under the gauge group must be real. The second term
can be written as
\begin{equation}
L_{Y}= -\frac{i}{2}\overline{\psi}\Gamma^{a}(\partial_{a}-
\frac{1}{2}f_{ibc}e^{i}_{\gamma}e^{\gamma}_{a}\Sigma^{bc}-
\frac{1}{2}G_{abc}\Sigma^{bc}- \phi_{a}) \psi \nonumber \\
=\frac{i}{2}\overline{\psi}\Gamma^{a}\nabla_{a}\psi+
\overline{\psi}V\psi ,
\end{equation}
where
\begin{eqnarray}
\nabla_{a}& =& - \partial_{a} +
\frac{1}{2}f_{ibc}e^{i}_{\gamma}e^{\gamma}_{a}\Sigma^{bc} + \phi_{a},\\
 V&=&\frac{i}{4}\Gamma^{a}G_{abc}\Sigma^{bc},
\end{eqnarray}
and we have used the full connection with torsion \cite{Batakis} given by
\begin{equation}
\theta_{c b}^{a} = -
f_{ib}^{a}e^{i}_{\alpha}e^{\alpha}_{c}-(D_{cb}^{a} +
\frac{1}{2}T_{c b}^{a}) = - f_{ib}^{a}e^{i}_{\alpha}e^{\alpha}_{c}-
G_{cb}^{a}
\end{equation}
with
\begin{equation}
D_{cb}^{a} = g^{ad}\frac{1}{2}[f_{db}^{e}g_{ec} + f_{cb}^{e}g_{de}
- f_{cd}^{e}g_{be}].
\end{equation}
A general choice of the torsion tensor which is S-invariant and
gives acceptable curvature two-form \cite{Review} is
\begin{equation}
T_{abc}= 2\tau(D_{abc} +D_{bca} - D_{cba}),
\end{equation}
where $\tau$ is a free parameter. In this case $T_{abc}$ is fully
antisymmetric and when the radii \cite{Review} are equal it is
proportional to $f_{abc}$. As we already have noticed, the CSDR
constraints tell us that $\partial_{a}\psi= 0$. Furthermore we can
consider the Lagrangian at the point $y=0$, due to its invariance
under S transformations, and as we showed $e^{i}_{\gamma}=0$ at
that point. Therefore eq.(21) becomes just $\nabla_{a}= \phi_{a}$
and the term $\frac{i}{2}\overline{\psi}\Gamma^{a}\nabla_{a}\psi $
in eq.(20) is exactly the Yukawa term.

Let us examine now the last term appearing in eq.(20). One can
show easily that the operator $V$ anticommutes with the
six-dimensional helicity operator \cite{Review}. Furthermore one
can show that $V$ commutes with the $T_{i}=
-\frac{1}{2}f_{ibc}\Sigma^{bc}$ ($T_{i}$ close the $R$-subalgebra
of $SO(6)$). In turn we can draw the conclusion, exploiting
Schur's lemma, that the non vanishing elements of $V$ are only
those which appear in the decomposition of both $SO(6)$ irreps $4$
and $\overline{4}$, e.g. the singlets. Since this term is a pure
geometric term, we reach the conclusion that the singlets in $4$
and $\overline{4}$ will acquire large geometrical masses, a fact
that has serious phenomenological implications. In supersymmetric
theories defined in higher dimensions, it means that the gauginos
obtained in four dimensions after dimensional reduction receive
superheavy masses, i.e. supersymmetry is broken at the
compactification scale. We note that for symmetric cosets the $V$
operator is absent because $f_{ab}^{c}$ are zero.

\section{Supersymmetry Breaking by Dimensional Reduction over Symmetric
Coset Spaces} Let us first consider the reduction of the
supersymmetric $E_{8}$ gauge theory over a symmetric coset space,
which is chosen to be the 6-sphere. Therefore $B=SO(7)/SO(6)$,
$D=10$ and we choose Weyl-Majorana fermions to belong to the
adjoint of G. The embedding of $R=SO(6)$ in $E_{8}$ is suggested
by the decomposition $$E_{8} \supset SO(6) \times SO(10) $$
\begin{equation}
248 = (15,1)+(1,45)+(6,10)+(4,16)+(\overline{4},\overline{16}).
\end{equation}
The $R=SO(6)$ content of the vector and spinor of $SO(7 )/SO(6)$
are $6$ and $4$ respectively. The condition for the anomaly
cancelation in four dimensions given in eq.(16)  is satisfied and
the four-dimensional gauge group is $H=C_{E_{8}}(SO(6))=SO(10)$.
The surviving scalars in four dimensions transform as a 10-plet
and the surviving fermions as a left handed $16$-plet under the
gauge group $SO(10)$. Therefore the four dimensional theory is an
anomaly free GUT with fermions in a multiplet which is appropriate
to describe a family of quark and leptons (including a right
handed neutrino), while any sign of the supersymmetry of the
original theory has dissappeared in the process of dimensional
reduction. Moreover we note that the Higgs content of the theory
is not appropriate to make the model phenomenologically
attractive. Concerning the spontaneous symmetry breaking of the
four dimensional theories resulting from CSDR the following
theorems hold :\\ (i) When $S/R$ is symmetric, the form of the
potential of the four-dimensional theory is such that it
necessarily leads to a spontaneous breakdown of $H$ \cite{Review},
and \\ (ii) When $S$ has an isomorphic image in $G$, the four
dimensional gauge group $H$ always breaks down spontaneously to a
subgroup $K$ which is the centralizer of $S$ in $G$
\cite{Vinet},\cite{Review}.

Therefore according to the above theorems in our model the
$SO(10)$ gauge symmetry is subject to a spontaneous breakdown,
while the final gauge group is $K= C_{E_{8}}(SO(7)) = SO(9) $.
Thus we see that the scalar field content could be appropriate for
the electroweak symmetry breaking but certainly cannot be
responsible for the first superstrong breaking at the GUT scale. A
mechanism for the superstrong breaking can be provided
\cite{Hosotani}, \cite{Review} and then the electroweak breaking
due to the Higgs fields is intimately connected to the
compactification scale.

\section{Supersymmetry Breaking by Dimensional Reduction over Non
Symmetric  Coset Spaces} Now we choose  $G=E_{8}$ and $B=
G_{2}/SU(3)$, which is a non-symmetric coset space. We use the
decomposition
\begin{eqnarray}
E_{8} &\supset& SU(3) \times E_{6}\nonumber \\
248 &=& (8,1) + (1,78) + (3,27) + (\overline{3},\overline{27}) ,
\end{eqnarray}
and we choose $SU(3)$ to be identified with $R$. The $R=SU(3)$
content  of $G_{2}/SU(3)$ vector and spinor is $ 3 + \overline{3}$
and $1+3$. The condition (16) for the cancellation of anomalies is
satisfied and the resulting four dimensional gauge group is $ H =
C_{E_{8}}(SU(3)) = E_{6}$, which contains fermion and scalar
fields transforming as 78, 27 and 27 respectively. Therefore we
obtain in four dimensions a supersymmetric anomaly free $E_{6}$
gauge theory with a vector superfield grouping gauge bosons and
fermions transforming according to the adjoint and a matter chiral
superfield grouping scalars and fermions in the fundamental of the
gauge group. Furthermore the very interesting feature of the CSDR
over the present coset space that we would like to stress here is
that the ${\cal N}=1$ supersymmetry of the four dimensional theory
is broken by soft terms. More precisely the scalar soft terms
appear in the potential of the theory and the gaugino masses come
from a geometric (torsion) term as already stated.

We proceed by calculating these terms. In order to determine the
potential we begin by examining the decomposition of the specific
$S$ under $R$, i.e.
\begin{eqnarray}
G_{2} &\supset& SU(3) \nonumber \\ 14 &=& 8+3+\overline{3}.
\end{eqnarray}
Corresponding to this decomposition we introduce the generators of
$G_{2}$
\begin{equation}
Q_{G_{2}} = \{ Q^{a},Q^{\rho},Q_{\rho}\},
\end{equation}
where $i=a,\ldots,8$ correspond  to the $8$ of $SU(3)$, while
$\rho = 1,2,3$ correspond to $3$ or $\overline{3}$. Then according
to the decomposition (28), the non trivial commutation relations
of the generators of $G_{2}$ are as follows
\begin{eqnarray}
\left[ Q^{a},Q^{b} \right] &=& 2i f^{abc}Q^{c},\nonumber \\
\left[ Q^{a},Q^{\rho} \right] &=&
-(\lambda^{a})^{\rho}_{\sigma}Q^{\sigma},\nonumber
\\
\left[ Q^{\rho},Q_{\sigma} \right] &=&
-(\lambda^{a})^{\rho}_{\sigma}Q^{a},\nonumber
\\
\left[ Q^{\rho},Q^{\sigma} \right] &=&
2\sqrt{\frac{2}{3}}\epsilon^{\rho\sigma\tau}Q_{\tau},
\end{eqnarray}
with normalization $$TrQ^{a}Q^{b}=2\delta^{ab},
TrQ^{\rho}Q_{\sigma}=2\delta^{\rho}_{\sigma} . $$

The potential of any theory reduced over $G_{2}/SU(3)$ can be written in
terms of the fields
\begin{equation}
\{ \phi_{i} , \phi^{\rho} ,\phi_{\rho} \},
\end{equation}
which correspond to the decomposition (28) of $G_{2}$ under
$SU(3)$. The $\phi_{i}$ are equal to the generators of the  $R$
subgroup. With the help of the commutation relations (30)  we find
that the potential of any theory reduced over $G_{2}/SU(3)$ is
given by \cite{Review} :
\begin{eqnarray}
V(\phi)=8+\frac{4}{3}Tr(\phi^{\rho}\phi_{\rho})
-\frac{1}{2}(\lambda^{i})^{\rho}_{\sigma}Tr(J_{i}[\phi_{\rho},\phi^{\sigma}])
+\sqrt{\frac{2}{3}}\epsilon^{\rho\sigma\tau}Tr(\phi_{\tau}[\phi_{\rho},\phi_{\sigma}])
\nonumber \\
 -\frac{1}{4} Tr([\phi_{\rho},\phi_{\sigma}][\phi^{\rho},\phi^{\sigma}] +
[\phi^{\rho},\phi_{\sigma}][\phi_{\rho},\phi^{\sigma}]).
\end{eqnarray}
Then to proceed with our specific choice $G=E_{8}$ we use the
embedding (27) of $R=SU(3)$ in $E_{8}$ and divide accordingly the
generators of $E_{8}$
\begin{equation}
Q_{E_{8}} = \{ Q^{a} , Q^{\alpha},Q^{i\rho},Q_{i\rho} \}
\end{equation}
with $a = 1,\ldots,8$,  $\alpha = 1,\ldots,78$, $i=1,\ldots,27$,
$\rho=1,2,3$. The normalization is $$TrQ^{a}Q^{b}=2\delta^{ab},\
TrQ^{\alpha}Q^{\beta}=12\delta^{\alpha\beta},\
TrQ^{i\rho}Q_{j\sigma}=2\delta^{i}_{j}\delta^{\rho}_{\sigma} . $$
The non trivial commutation relations of the generators of $E_{8}$
according to the decomposition (27) are the following :
\begin{eqnarray}
\left[ Q^{a},Q^{b} \right]&=&2if^{abc}Q^{c},\nonumber \\
\left[ Q^{\alpha},Q^{\beta}
\right]&=&2ig^{\alpha\beta\gamma}Q^{\gamma},\nonumber\\
\left[ Q^{a},Q^{i\rho}
\right]&=&-(\lambda^{\alpha})^{\rho}_{\sigma}\delta^{i}_{j}Q^{j\sigma},
\nonumber\\
\left[ Q^{i\rho},Q^{j\sigma} \right]&=&\frac{1}{\sqrt{6}}
\epsilon^{\rho\sigma\tau}d^{ijk}Q_{k\tau},\nonumber\\
\left[ Q^{i\rho},Q_{j\sigma}
\right]&=&-(\lambda^{a})^{\rho}_{\sigma}\delta^{i}_{j}Q^{a}+
\frac{1}{6}\delta^{\rho}_{\sigma}(G^{\alpha})^{i}_{j}Q^{\alpha},\nonumber\\
\left[ Q^{\alpha},Q^{i\rho}
\right]&=&(G^{\alpha})^{i}_{j}\delta^{\rho}_{\sigma}Q^{j\sigma} ,
\end{eqnarray}
where $d^{ijk}$, the symmetric invariant $E_{6}$ tensor, and
$(G^{\alpha})^{i}_{j}$ are defined in ref.
 \cite{Kephart}.

Next we would like to solve the constraints (19) and examine the
resulting four dimensional potential in terms of the unconstrained
scalar fields $\beta$. The solutions of the constraints in terms
of the genuine Higgs fields are
\begin{equation}
\phi^{a}=Q^{a},  \phi_{\rho}=\beta^{i}Q_{i\rho},
\phi^{\rho}=\beta_{i}Q^{i\rho}.
\end{equation}
In turn we can express the Higgs potential in terms of the genuine Higgs
field $\beta$ and we find
\begin{equation}
V(\beta)= 8- \frac{40}{3}\beta^{2} - 4d_{ijk}\beta^{i}\beta^{j}\beta^{k}
+
\beta^{i}\beta^{j}d_{ijk}d^{klm}\beta_{l}\beta_{m}+
\frac{11}{4}\sum_{\alpha}\beta^{i}(G^{\alpha})_{i}^{j}
\beta_{j}\beta^{k}(G^{\alpha})_{k}^{l}\beta_{l}.
\end{equation}
From the potential given in eq.(36) we can read directly the F-,
D- and scalar soft terms which break softly the supersymmetric
theory obtained by CSDR over $G_{2}/SU(3)$. The F-terms are
obtained from the superpotential :
\begin{equation}
{\cal W} (B) =\frac{1}{3}d_{ijk}B^{i}B^{j}B^{k},
\end{equation}
where $B$ is the chiral superfield whose scalar component is
$\beta$. Let us note that the superpotential could also be
identified from the relevant Yukawa terms of the fermionic part of
the Lagrangian. Correspondingly the D-terms are
\begin{equation}
D^{\alpha} =\sqrt{\frac{11}{2}}\beta^{i}(G^{\alpha})^{j}_{i}\beta_{j}.
\end{equation}
The terms in the potential $V(\beta)$ given in eq.(36) that do not
result from the F- and D-terms belong to the soft supersymmetry
part of the Lagrangian. These terms are the following
\begin{equation}
L_{SSB}=-\frac{40}{3}\beta^{2}-4d_{ijk}\beta^{i}\beta^{j}\beta^{k}.
\end{equation}

Finally in order to determine the gaugino mass we calculate the V
operator given in eq.(22). Using eq.(24) we find that
\begin{equation}
D_{abc}=\frac{1}{2}f_{abc}
\end{equation}
and
in turn the $G_{abc}=D_{abc}+\frac{1}{2}T_{abc}$ is
\begin{equation}
G_{abc}=\frac{1}{2}(1-3\tau)f_{abc}.
\end{equation}
In order to obtain the previous
results the
most general $G_{2}$ invariant metric on $G_{2}/SU(3)$ was used which is
$g_{ab}=r\delta_{ab}$.\\ In addition we  need the gamma matrices. In ten
dimensions
we have
$ [\Gamma^{\mu},\Gamma^{\nu}] = 2 \eta^{\mu\nu}$ with
$\Gamma^{\mu}= \gamma^{\mu}\otimes I_{8}$ and
$[\Gamma^{a},\Gamma^{b}] = -2g^{ab}$, where
$$ \Gamma^{a} =
\frac{1}{\sqrt{r}}\gamma_{5}\otimes\left[\begin{array}{cc}0&\overline{\gamma}^{a}\\

\overline{\gamma}^{a}&0 \end{array} \right] $$
with $a=1,2,3,5,6$ and
$$\Gamma^{4}=\frac{1}{\sqrt{r}}\gamma_{5}\otimes\left[\begin{array}{cc}0&iI_{4}\\

iI_{4}&0 \end{array} \right]. $$ The $\overline{\gamma}^{a}$
matrices are given by $
\overline{\gamma}^{1}=\sigma^{1}\otimes\sigma^{2}$ ,
$\overline{\gamma}^{2}=\sigma^{2}\otimes\sigma^{2}$,
$\overline{\gamma}^{3}=-I_{2}\otimes\sigma^{3}$,
$\overline{\gamma}^{5}=\sigma^{3}\otimes\sigma^{2}$,
$\overline{\gamma}^{6}=-I_{2}\otimes\sigma^{1}$. Using these
matrices we calculate
$\Sigma^{ab}=\frac{1}{4}[\Gamma^{a},\Gamma^{b}]$. We find that the
gauginos acquire a geometrical mass
$$(1-3\tau)\frac{6}{\sqrt{3}}.$$

Therefore by reduction of a ten dimensional supersymmetric $E_{8}$
gauge theory over the non symmetric coset space $G_{2}/SU(3)$, we
obtain in four dimensions a Lagrangian describing a supersymmetric
$E_{6}$ gauge theory as well as its full soft supersymmetry
breaking sector. We note that the trilinear and scalar masses soft
breaking terms depend only on the radius of the $G_{2}/SU(3)$
while the gaugino has an additional dependence on the possible
torsion of the coset space and can be therefore further adjusted.
\section{Conclusion}
 In conclusion, we have presented examples of dimensional reduction of a
ten-dimensional supersymmetric gauge theory over six-dimensional
coset spaces. Our aim was to show, extending and completing
previous observations \cite{Chapline}, \cite{Kapetanakis},
\cite{Review}, that the CSDR of a supersymmetric gauge theory over
non symmetric coset spaces results to a softly broken
supersymmetric gauge theory in four dimensions. Thus CSDR provides
us with a very interesting mechanism for supersymmetry breaking. A
more complete presentation of similar results involving other
six-dimensional coset spaces will be given elsewhere due to lack
of space here. For completeness we have briefly included in the
present examination the CSDR over symmetric coset spaces, which
was known that leads to complete breaking  of supersymmetry, and
can be very useful in attempts to extract physics from p-branes.

We would like to thank P.~Forgacs, A.~Kehagias, C.~Kounnas,
G.~Koutsoumbas  and D.~Luest for useful discussions and reading
the manuscript.

\end{document}